\documentstyle[11pt,aasms4]{article}

\received{April 6 2000}
\accepted{June 3 2000}
\begin{document}

\title{On the radial density profile of intracluster gas tracing
the isothermal dark halo with a finite core}

\author{Xiang-Ping Wu and Yan-Jie Xue}

\affil{Beijing Astronomical Observatory and 
National Astronomical Observatories,
Chinese Academy of Sciences, Beijing 100012; China}

\begin{abstract}
The cusped NFW universal density profile suggested by typical CDM models
has been challenged in recent years by the discoveries of
the soft cores with finite central density for a broad range of masses from
dwarf galaxies to clusters of galaxies. It is thus desirable that 
a new, analytic model would instead become available for virialized dark halos.
One promising candidate is probably the empirical density profile proposed by
Burkert (1995), which resembles an isothermal profile 
with a constant core in the inner region and matches the NFW 
profile at large radii. Meanwhile, such a revised dark halo (RDH) profile 
has turned out a great success on galactic scales. 
This stimulates us to  apply the RDH profile to more massive systems 
like clusters of galaxies. In this paper we have made an attempt to derive  
the radial density profile of intracluster gas from the RDH profile under
the isothermal and hydrostatic equilibrium hypotheses, 
and compare it with those revealed by X-ray observations and inferred from 
the NFW profile. It is shown that the RDH predicted gas density can be well
represented by the conventional $\beta$ model with a typical $\beta$ parameter 
of $\beta\sim0.7$--$0.9$. 
Alternatively, fitting the theoretically predicted X-ray 
surface brightness profile to an ensemble of 45 X-ray clusters observed 
by ROSAT, we find that the RDH and NFW profiles become to be almost 
indistinguishable from each other, and their characteristic 
density and scale length parameters are strongly correlated.
Yet, unlike the NFW model, the RDH profile can allow us to work out
straightforwardly the central dark matter density from X-ray measurements 
of the surface brightness and temperature of clusters. It appears that
the resulted central densities of the 45 clusters have an average value of 
$\langle\rho_0\rangle\approx0.01$ M$_{\odot}$ pc$^{-3}$, 
in agreement with the result estimated on galactic scales,
which reinforces the claim for the presence of the soft halo cores 
over the entire mass range.
\end{abstract}


\section{Introduction}

For decades many efforts have been made towards the understanding of the 
radial density profile of intracluster gas from the well-motivated
physical mechanism, in an attempt to recover the 
empirical $\beta$ model (Cavaliere \& Fusco-Femiano 1976) which
fits nicely the X-ray observed surface brightness distribution of clusters.
Assuming both galaxies and gas are the tracers of the shape and depth of
a common gravitational potential of a cluster, 
Cavaliere \& Fusco-Femiano (1976, 1978) obtained
an analytic gas density profile resembling the $\beta$ model in shape
if the King model is used to represent the galaxy number density profile 
in the inner  cluster region. In recent years, 
the rapid progress of numerical simulating techniques 
has permitted the reconstruction of the dark matter halos with 
an unprecedented resolution,  ranging from galactic scales of $\sim 1$ kpc
to large-scale structures of $\sim10$ Mpc. This leads one to view the
issue at a different angle: Given the gravitational potential wells
defined by the dark halos, how is the intracluster gas distributed in
clusters if the hydrostatic equilibrium  between the
gas and the underlying gravitational potentials has been built up ?
In particular, much attention has been paid to the issue of how to
reconstruct the radial gas density and temperature profiles if the
dark halos follow the so-called universal density profile 
(Navarro, Frenk \& White 1995; NFW) suggested 
by  high-resolution simulations within the framework of 
typical CDM models such as SCDM, LCDM and $\Lambda$CDM 
(Makino, Sasaki \& Suto 1998; Suto, Sasaki \& Makino 1998; 
Yoshikawa \& Suto 1999; Wu \& Chiueh 2000). 
It turns out that there is indeed  a striking similarity between
the predicted X-ray surface brightness profiles of 
clusters and the conventional $\beta$ model.
This has stimulated several authors to apply the NFW predicted X-ray
surface brightness profiles to the observed ones for an ensemble
of X-ray clusters (Makino \& Asano 1999; Ettori \& Fabian 1999; 
Wu \& Xue 2000; Wu 2000).

Yet, besides its uncomfortable singularity at $r=0$,
the cusped NFW profile has been shown to be
in conflict with observations (e.g. Tyson, Kochanski \& Dell'Antonio 1998; 
Navarro \& Steinmetz 2000; Firmani et al. 2000; and references therein). 
In fact, it has been noticed
that, while the NFW profile yields a gas density profile close to the 
empirical $\beta$ model, the core radius predicted by the cusped
density profile may be smaller than the actually observed one
(Makino et al. 1998). Moreover, the NFW profile leads to
an increasing gas temperature towards cluster centers (Wu \& Chiueh 2000), 
in contrast with the presence of the cold gas components detected
very often inside the X-ray cores. Motivated by the soft inner matter
distributions of the dark halos revealed observationally
from dwarf galaxies to rich clusters,
Spergel \& Steinhardt (2000) recently proposed  
that the CDM particles are self-interacting. As a result, the collisional
CDM particles, in a similar way to the baryonic particles, will produce 
the less centrally concentrated structures. This weakly self-interacting
CDM model has soon attracted many investigations, among which the numerical
simulations of structure formation based on some simple physical
consideration of the self-interacting CDM particles have 
successfully provided a scenario that is essentially consistent 
with the existing observations
(Hannestad 1999; Burkert 2000 and references therein).

As a natural extension, one may address the following question:
Does there exist a similar analytic expression to the NFW universal
density profile that one can use to approximately describe the matter
distribution of the virialized dark halos 
resulted from the weakly self-interacting CDM model ?  
A conclusive answer to such a question seems not easy:
Unlike the standard CDM particles, for which there is no need to
consider the interaction between particles except their gravity, 
the collisional, warm CDM particles contain an unknown parameter, 
the cross-section. Consequently, even with the help of  
high-resolution simulations, it is still hard to completely 
determine the final configurations of dark halos. 
Under present circumstances,  one promising  candidate is probably 
the empirical density profile suggested by Burkert (1995): 
%1
\begin{equation}
\rho_{DM}(r)=\frac{\rho_0r_0^3}{(r+r_0)(r^2+r_0^2)},
\end{equation}
where $\rho_0$ and $r_0$ are the central density and the scale length,
respectively. This revised dark halo (RDH) density law
resembles an isothermal profile
in the inner region with a constant core $r_0$, while in the outer region 
the mass profile diverges logarithmically with $r$, in agreement with 
the NFW profile. So, such a density profile does have the desired 
properties at the central and outermost regions. 
In particular, it fits fairly well the dark matter distributions of 
dwarf galaxies revealed by both the rotation curves and the numerical
simulations of evolution of halos consisting of weakly self-interacting
CDM particles (Burkert 2000; Salucci \& Burkert 2000).

Motivated by the apparent success of the RDH profile 
of eq.(1) on galactic scales and the possible existence of 
weak interaction between CDM particles, in this paper we would like 
to apply the RDH profile to more massive systems like clusters of galaxies.
We would like to demonstrate  the radial density profile of intracluster gas 
tracing the gravitational potential defined
by the RDH profile, and compare the expected X-ray surface brightness 
with the X-ray observations and other models (e.g. the conventional $\beta$ 
model and that predicted by the NFW profile), although we have an 
intuition that the RDH profile, as a combination of the 
NFW profile and the $\beta$ model,  would provide an essentially
similar result. This study will nevertheless
constitute an important test for the universality of 
the RDH profile as the virialized dark halos over the entire mass range,
and will also allow us to examine whether there are any common properties 
in the NFW and RDH profiles. Eventually, it is hoped that this study 
will be useful for a conclusive answer to the question as to  
whether the RDH profile can be used 
to replace the role of the NFW profile for the virialized dark halos,
if CDM particles are indeed weakly self-interacting.

\section{Density profile of intracluster gas}

\subsection{Self-gravity of the gas: excluded}

If the dark halo of a cluster follows the RDH profile described by eq.(1),
the total dark matter enclosed within radius $r$ is
%2,3
\begin{eqnarray}
M_{DM}(x)=4\pi \rho_0 r_0^3 \tilde{m}(x);\\
\tilde{m}(x)=\frac{1}{2} \left[\ln(1+x) + \frac{1}{2}\ln{(1+x^2)}
             -\arctan(x)\right],
\end{eqnarray}
where $x=r/r_0$. Assuming that the intracluster gas is isothermal (with
a temperature of $T$) and in hydrostatic equilibrium with 
the underlying gravitational potential dominated by $M_{DM}$, we have
%4
\begin{equation}
\frac{GM_{DM}(x)}{x^2}=-\frac{kTr_0}{\mu m_p n_{gas}(x)} 
        \frac{dn_{gas}(x)}{dx},
\end{equation}
in which $n_{gas}(x)$ is the gas number density and $\mu$ is the average 
molecular weight. Note that  
we have neglected the self-gravity of the gas for the moment.
A straightforward computation yields an analytic form of gas density
profile:
%5
\begin{equation}
\frac{n_{gas}(x)}{n_{gas}(0)}=\left[e^{-(1+\frac{1}{x})\arctan x}
                       (1+x)^{(1+\frac{1}{x})}
		       (1+x^2)^{\frac{1}{2}(\frac{1}{x}-1)}\right]
                 ^{\frac{\alpha_0}{2}},
\end{equation}
where
%6
\begin{equation}
\alpha_0=\frac{4\pi G\mu m_p\rho_0 r_0^2}{kT}.
\end{equation}
In order to avoid the divergence of the resulting X-ray surface brightness 
to be discussed below because of  
$n_{gas}(\infty)=n_{gas}(0)e^{-\pi\alpha_0/4}$, 
in a similar way to the
treatment of the NFW predicted gas density profile  (see Wu 2000),  
we introduce a normalized, background subtracted gas number density 
$\tilde{n}_{gas}(x)\equiv [n_{gas}(x)-n_{gas}(\infty)]/
[n_{gas}(0)-n_{gas}(\infty)]$, which reads
%7
\begin{eqnarray}
\tilde{n}_{gas}(x)=
		\frac{1}{e^{\frac{\pi\alpha_0}{4}}-1}
                                                             \nonumber \\
           \left\{     \left[e^{\frac{\pi}{2}-(1+\frac{1}{x})\arctan x}
                       (1+x)^{(1+\frac{1}{x})}
		       (1+x^2)^{\frac{1}{2}(\frac{1}{x}-1)}\right]
                 ^{\frac{\alpha_0}{2}} -1 \right\}.
\end{eqnarray}
Another way to deal with the non-zero background gas density 
predicted by the dark halo model is to truncate the cluster
at a certain radius (e.g. the virial radius) as adopted by
Makino et al. (1998).  It is indeed unfortunate that one has to
add an arbitrary, unphysical constraint on the gas density profile to
ensure the convergence of the X-ray surface brightness.

In Fig.1 we demonstrate the radial profiles of the scaled gas density 
$\tilde{n}_{gas}(x)$ for typical clusters with 
$\alpha_0=5$, $10$ and $20$.  
A glimpse of Fig.1 seems to suggest that all the
resulted profiles of $\tilde{n}_{gas}(x)$  
resemble the conventional $\beta$ models in shape. We then overlap 
the $\beta$ model, 
$\tilde{n}^*_{gas}(x)=[1+(x/x_c)^2]^{-3\beta/2}$, to
each curve with ($\beta$, $r_c/r_0$) $=$ (0.40, 0.86), (0.74, 0.85) and
(1.56, 0.92) for $\alpha_0=$5, 10 and 20, respectively,
where $x_c=r_c/r_0$ is the scaled core radius. 
Yet, like the actual fitting of the $\beta$ model to the X-ray observed 
surface brightness profile,
the best-fit $\beta$ parameters depend also on the extension of 
the fitting regions. For a typical cluster of $\alpha_0\approx10$,
the best-fit $\beta$ value over a region out to $10r_c$ is 
$\beta\approx0.7$, in good agreement with X-ray 
observations. In particular, the gas core radius takes roughly the
value of $r_0$.
Also plotted in Fig.1 are the corresponding gas densities predicted by
the NFW profile with $\alpha=\alpha_0$ (Makino et al. 1998; Wu 2000)
%8
\begin{equation}
\tilde{n}_{gas}(x)=
                 \frac{(1+x)^{\alpha/x}-1}{e^{\alpha}-1},
\end{equation}
where $x=r/r_s$ and $\alpha=4\pi G\mu m_p\rho_s r_s^2/kT$.
A visual examination of Fig.1 reveals that 
there are some differences between the expected 
radial variation of intracluster gas
from the RDH profile and that from the NFW  profile, which
are reflected not only by the significantly different scale lengths  
but also by the different shape. Indeed, it is unlikely that 
the two types of density profiles can be made to be identical simply 
by a horizontal replacement. Whether these differences are significantly 
important will be discussed when the two density profiles are
both applicable to an ensemble of X-ray clusters (section 3).

\placefigure{fig1}

\subsection{Self-gravity of the gas: included}

The total mass in gas within radius $x$ is simply
%9
\begin{equation}
M_{gas}(x)= 4\pi\mu m_p r_0^3 \int_0^{x} n_{gas}(x) x^2 dx.
\end{equation}
When the self-gravity of the gas is included, the hydrostatic equation
becomes
%10
\begin{equation}
\frac{G[M_{DM}(x)+M_{gas}(x)]}{x^2}
=-\frac{kT r_0}{\mu m_p n_{gas}(x)} \frac{dn_{gas}(x)}{dx}.
\end{equation}
If we introduce the volume-averaged (gas) baryon fraction $f_b(x)$ as a new 
variable:
%11
\begin{equation}
f_b(x)=\frac{M_{gas}(x)}{M_{DM}(x)+M_{gas}(x)},
\end{equation}
we can obtain the following two first-order differential equations
%12,13
\begin{eqnarray}
\frac{d\bar{n}_{gas}}{dx}=-\alpha_0
                    \frac{\tilde{m}\bar{n}_{gas}}{(1-f_b)x^2};\\
\frac{df_b}{dx}=\frac{(1-f_b)[b(1-f_b)\bar{n}_{gas}-f_b\tilde{\rho}_{DM}]x^2}
                {\tilde{m}},
\end{eqnarray}
where $\bar{n}_{gas}={n}_{gas}(x)/{n}_{gas}(0)$, 
$b=\mu m_p n_{gas}(0)/\rho_0$ and $\tilde{\rho}_{DM}=\rho_{DM}/\rho_0=
(1+x)^{-1}(1+x^2)^{-1}$. We need to specify the boundary conditions
in order to solve the above equations. The first
condition is obviously
%14
\begin{equation}
\bar{n}_{gas}(0)=1.
\end{equation}
Following the recent work of Wu \& Chiueh (2000), 
we choose the boundary condition of $f_b(x)$ such that the baryon fraction
within the virial radius $r_{vir}$ (or $c=r_{vir}/r_0$) should 
asymptotically approach the universal value $f_{b,BBN}=\Omega_b/\Omega_M$,
where $\Omega_b$ and $\Omega_M$ are, respectively,
the baryon and total mass densities of the Universe in units of the 
critical density $\rho_{c}$ for closure, namely, 
%15,16
\begin{eqnarray}
f_b(c)=f_{b,BBN};\\
\frac{df_b}{dx}\left|_{x=c}=0. \right.
\end{eqnarray}
Here the scaled virial radius or the so-called concentration parameter $c$ 
is defined by
%17
\begin{equation}
M_{DM}(c)=\frac{4\pi}{3}r_0^3\rho_{c} c^3\Delta_c,
\end{equation}
or
%18
\begin{equation}
\frac{\tilde{m}(c)}{c^3}=\frac{\Delta_c}{3}\frac{1}{\delta_c},
\end{equation}
in which $\delta_c=\rho_0/\rho_{c}$, and  $\Delta_c$ represents 
the overdensity parameter of dark matter with respect to the
average background value $\rho_c$ and
will be taken to be $200$ in our computation below.

The free parameters involved in eqs.(12) and (13) and the 
boundary conditions are $\alpha_0$, $b$, $f_{b,BBN}$ and
$c$ or $\delta_c$. However, there are only two independent parameters
with the restrictions of eqs.(15) and (16), which we choose to be
$\delta_c$ and $f_{b,BBN}$ below. We will perform 
the numerical searches for the solutions of eqs.(12) and (13) 
under the boundary conditions of eqs.(14)--(16). 
Technically, we search for the 
solutions over a two-parameter space ($\alpha_0$, $b$) by iterations 
until the boundary conditions are satisfied,
which enables us to work out the radial profiles
of gas density and baryon fraction, together with a unique determination
of the parameters $\alpha_0$ and $b$.  In Fig.2 we demonstrate a set
of solutions for two typical choices of $f_{b,BBN}$ and $\delta_c$: 
$f_{b,BBN}=(0.05$, $0.1)$ and $\delta_c=(10^4$, $10^5)$, and
the resulting $\alpha_0$ and $b$ values are listed in Table 1.
It appears that the derived density profiles of the baryon fraction exhibit 
no dramatic variation over the whole clusters: For a small value of 
$\delta_c\sim10^4$, $f_b$ increases slightly with outward radius and 
eventually matches the background value at virial radius, while 
for a large $\delta_c\sim10^5$,  $f_b$ would reach a maximum  before
approaching asymptotically the universal value. 
The parameter $\alpha_0$ also acts roughly like a constant ($\sim12$) 
even if the characteristic density $\delta_c$ changes by a decade,
and the parameter $b$ turns out to be an increasing function of  
$f_{b,BBN}$, which arises simply from $b\propto \mu m_p n_{gas}(0)$.
Now, we concentrate on the derived profiles of gas density.
All the predicted density curves of intracluster gas shown in Fig.2 
seem to well resemble the $\beta$ models, which is illustrated by our 
superimposed $\beta$ model onto each of the derived density profiles. 
The most remarkable feature is that 
the corresponding values of $\beta$ and core radius have all fallen 
into very narrow ranges of $0.87<\beta<0.98$ and $0.65<r_c/r_0<0.88$ 
for our choices of the universal baryon fraction 
$f_{b,BBN}$ and the characteristic density $\delta_c$ for typical clusters 
(see Table 1).  An increase of $\delta_c$ up to $10^{6}$, the roughly 
largest density for clusters (see next section),   only leads to 
a minor modification to these limits. This indicates that
the isothermal gas in different clusters 
should essentially follow a similar distribution, i.e., 
the observationally determined $\beta$ among different clusters should 
not show a large scatter around the mean value $\beta\sim0.9$.
Note that the gas core radius $r_c$ may vary substantially because of 
the different $r_0$ for different clusters.
While the predicted density profile of intracluster gas demonstrates
a property basically consistent with what has been known for X-ray 
clusters, the theoretically expected $\beta$ parameter is likely to
slightly exceed the presently determined one from X-ray observation,
$\beta\approx0.7$. 
The former arises mainly from our restriction that the baryon fraction 
should asymptotically match the background value. Previous 
studies with this constraint have also arrived at a similar conclusion:
the $\beta$ value in the $\beta$ model for intracluster gas is required 
to be larger than a certain low-limit (Wu \& Chiueh 2000).
Actually, it is not impossible that the presently fitted $\beta$ parameters
based on the X-ray observed surface brightness profiles of clusters  
are biased low because of the influence of the cooling-flows and the
small fitting regions. Excluding the cooling flows or adopting a
double $\beta$ model fit may moderately raise the observationally determined
$\beta$ values, giving rise to $\beta\approx0.7$--$0.8$ 
(Vikhlinin et al. 1999; Xue \& Wu 2000). Of course, the present computation
has been made within the framework of isothermality, and the above
prediction cannot be taken too literally unless the non-isothermal gas 
is included. 

\placefigure{fig2}

 \begin{deluxetable}{ccccccccc}
 \tablewidth{30pc}
% \scriptsize
 \tablecaption{Numerical results for typical clusters}
 \tablehead{
\colhead{$f_{b,BBN}$} & \colhead{$\delta_c$} & \colhead{$c$} & 
\colhead{$\alpha_0$}  & \colhead{$b$} & \colhead{$\beta$} & 
\colhead{$r_c/r_0$}   & \colhead{$\beta^*$} & \colhead{$(r_c/r_0)^*$} }
 \startdata
0.05 & $10^4$ &  5.48  & 12.0 & 0.05 & 0.867 & 0.705 & 0.90 & 0.66 \nl
0.05 & $10^5$ &  14.2  & 13.7 & 0.09 & 0.974 & 0.681 & 0.83 & 0.68 \nl
0.10 & $10^4$ &  5.48  & 10.8 & 0.09 & 0.927 & 0.884 & 0.90 & 0.63 \nl
0.10 & $10^5$ &  14.2  & 12.8 & 0.23 & 0.981 & 0.649 & 0.83 & 0.68 
\tablenotetext{*}{The $\beta$ model fit to the approximate solution
eq.(7) by assuming $f_b=f_{b,BBN}$ and replacing $\alpha_0$ by 
$\alpha_0/(1-f_{b,BBN})$.}

\enddata
 \end{deluxetable}

Considering the fact that $f_b$ exhibits only a minor variation over
the whole cluster, and also for illustrating the effect of the 
self-gravity of the gas, we can provide an approximate and analytic 
form of the gas density by taking $f_b=f_{b,BBN}$ in eqs.(12) and (13).
Consequently, eq.(12) reduces to eq.(4) and the gas density is given
by the analytic expression eq.(7),  in which the parameter $\alpha_0$ 
is now replaced by  $\alpha_0^*=\alpha_0/(1-f_b)$.  In Table 1, we list
the $\beta$ parameters and core radii in the $\beta$ model fits to 
our approximate solutions using the same input
values of $f_{b,BBN}$ and $\delta_c$ (or $\alpha_0$). In the case of the low 
density $\delta_c=10^4$, the agreement between the 
exact and approximate solutions is fairly good, while the approximate
solutions seem to underestimate the $\beta$ values by $\sim0.15$
for  $\delta_c=10^5$. The latter is partially  due to the fact that 
the fitting of the $\beta$ model is made over rather a large region out 
to $x=c=14.2$, where a shallower density profile occurs 
according to the RDH profile (see Fig.1). 
Another reason is that in the case of $\delta_c=10^4$, 
the baryon fraction remains roughly unchanged across the clusters 
with a relative variation rate of $|f_b(c)-f_b(0)|/f_{b,BBN}<20\%$,   
in comparison with  $|f_b(c)-f_b(0)|/f_{b,BBN}\sim100\%$ for 
$\delta_c=10^5$. Namely, the small $\delta_c$ clusters seem to
meet more easily the condition for the approximate solutions ($f_b=f_{b,BBN}$)
than the large $\delta_c$ clusters do.  Nevertheless, we conclude that
the derived gas distributions with and without the inclusion of the
self-gravity of the gas do not show very significant difference, and
in general, the self-gravity of the gas only leads to a slightly 
steeper gas density profile as a result of the increase of 
the underlying gravitational mass. This is consistent with the similar study 
for the NFW profile (Suto et al. 1998).

\section{Application to X-ray clusters}

In this section we conduct a comparison between our derived gas density 
profile and X-ray observations, which are linked up through the X-ray 
surface brightness profiles of clusters, $S_x$. 
In the scenario of the optically thin, isothermal plasma emission, 
%19
\begin{equation}
S_x(x)\propto \int_x^{\infty} \tilde{n}^2_{gas} d\ell,
\end{equation}
where the integral is performed along the line of sight $\ell$.
Here we take the analytic form of the gas density eq.(7) 
and neglect the contribution of 
the gas self-gravity. We intend to fit our theoretically expected
X-ray surface brightness profile eq.(19) to an ensemble of the X-ray
observed surface brightness profiles of clusters, which will allow  
us to determine the characteristic parameters, $\rho_0$ and $r_0$,
for the RDH profile and compare them with the results from other models,
e.g. the $\beta$ model and the NFW profile.

We use the ROSAT PSPC observed surface brightness profiles of 45 nearby 
clusters compiled by Mohr, Mathiesen \& Evrard (1999, MME). 
Several models have been already tested with this sample, 
such as the $\beta$ model, the double $\beta$ model and 
the NFW predicted density profile (MME; Wu \& Xue 2000; Xue \& Wu 2000).
In a similar way to our previous analysis for the NFW profile 
(Wu \& Xue 2000), we perform the $\chi^2$-fit to get the best-fit 
parameters $\alpha_0$ and $r_0$, in which we keep the same outer radii
of the fitting regions as those defined by MME. 
In order to examine how our results  are affected by the presence of 
the cooling flows in some clusters, we perform our fittings  by using
the entire data points of $S_x$ (model A) and 
excising the central region of 0.05 Mpc in each cluster (model B),
respectively. For the latter the reason that we adopt
the same inner radius for all the clusters is to  
guarantee the uniformity of the excision (Markevitch 1998).
We then compute $\rho_0$ in terms
of eq.(6) by  taking the X-ray temperature data from the literature
(see Wu, Xue \& Fang 1999; and references therein).
For majority of the clusters we use the cooling flow corrected 
temperature data by White (2000). 
In Fig.3 we illustrate a typical example of the observed and our fitted
surface brightness profiles (model A) for cluster A3158. 
For comparison, we have
also plotted the results of the $\beta$ model and the NFW profile. 
Essentially, these three models provide more or less an equal goodness of 
fit to the observed data with $\chi^2_{\nu}=1.12$, $1.24$ and $1.10$ for
the $\beta$ model, the NFW and RDH profiles, respectively. 
From the fitting of the X-ray surface brightness profile alone, 
it may be hard to reject any of these models.
In Table 2 we list the best-fit values of  $\rho_0$ and $r_0$ for
the 45 MME clusters by model A, 
together with the results for the $\beta$ model and the 
NFW profile, in which all the quoted errors are $68\%$ confidence limits.
The Hubble constant is taken to be $H_0=50$ km s$^{-1}$ Mpc$^{-1}$. 
When the central regions of 0.05 Mpc are excised in the fittings,
the best-fit parameters for all the three models are only
moderately affected. However, this can significantly improve 
the goodness of the fittings characterized by the reduced $\chi^2_{\nu}$:
After the excision of the central regions (model B),   
the fractions of clusters with  $\chi^2_{\nu}\leq2$ increase from 
(10/45, 14/45, 7/45) to (26/45, 25/45, 22/45)  for  
($\beta$, NFW, RDH) models, respectively.
Recall that the similar fraction of 26/45 is found by  MME 
for the single $\beta$ and double $\beta$ model fittings. 
It appears that about half of the clusters cannot be well fitted by
any of these models even if the central regions are excised.

% \begin{deluxetable}{llllllllll}
% \tablewidth{40pc}
% \scriptsize
% \tablecaption{cluster sample} 
% \tablehead{
% \colhead{cluster} &
% \colhead{$T$ (keV)} &
% \colhead{$r{_c}$ (Mpc)} & 
% \colhead{$\beta$} & 
% \colhead{$\alpha$} & 
% \colhead{$r{_s}$ (Mpc)} & 
% \colhead{$\rho{_s}$ ($10{^4}\rho{_c}$)} & 
% \colhead{$\alpha{_0}$} & 
% \colhead{$r{_0}$ (Mpc)} &  
% \colhead{$\rho{_0}$ ($10{^4}\rho{_c}$)} } 
% \startdata
% & & & & &  see table 2 & & & & \nl
% \enddata
% \end{deluxetable}

\placefigure{fig3}

\subsection{RDH vs. $\beta$ model}

We first analyze the possible link between the RDH profile and
the $\beta$ model. For this purpose we plot in Fig.4  the best-fit 
$\alpha_0$ vs. $\beta$ and $r_0$ vs. $r_c$ for the 45 MME clusters
obtained by model A. 
It is immediate that there exist strong correlations between 
the scale and slope parameters in the two models. We fit these correlations 
to a power-law function, which reads
%20,21
\begin{eqnarray}
\alpha_0=10^{1.08\pm0.01}\beta^{0.67\pm0.06}, & \;{\rm (A)};\\
\alpha_0=10^{1.10\pm0.01}\beta^{0.86\pm0.04}, & \;{\rm (B)},
\end{eqnarray}
and
%22,23
\begin{eqnarray}
r_0=10^{-0.071\pm0.015} r_c^{0.75\pm0.02}, & \;{\rm (A)};\\
r_0=10^{-0.033\pm0.012} r_c^{0.83\pm0.01}, & \;{\rm (B)}.
\end{eqnarray}
Meanwhile, applying a linear fit to the data set yields
%24,25
\begin{eqnarray}
\alpha_0=(14.11\pm1.09)\beta , & \;{\rm (A)};\\
\alpha_0=(13.44\pm0.55)\beta, & \;{\rm (B)}, 
\end{eqnarray}
and
%26,27
\begin{eqnarray}
r_0=(1.46\pm0.41) r_c, & \;{\rm (A)};\\
r_0=(1.30\pm0.18) r_c, & \;{\rm (B)}.
\end{eqnarray}
It appears that the resultant relationships with and without
the excision of the central regions in the fits of the X-ray observed
surface brightness profiles are roughly consistent with each other.

\placefigure{fig4}

\subsection{NFW vs. $\beta$ model}

For comparison we display in Fig.5 the correlations between the corresponding
parameters in the NFW profile and the $\beta$ model. 
The best-fit $\alpha$-$\beta$ and $r_s$-$r_c$ relations 
using all the 45 data points are
%28,29
\begin{eqnarray}
\alpha=10^{1.28\pm0.02}\beta^{1.46\pm0.08}, & \;{\rm (A)};\\
\alpha=10^{1.36\pm0.02}\beta^{1.81\pm0.08}, & \;{\rm (B)},
\end{eqnarray}
and
%30,31
\begin{eqnarray}
r_s=10^{0.64\pm0.03} r_c^{1.03\pm0.03}, & \;{\rm (A)};\\
r_s=10^{0.77\pm0.02} r_c^{1.27\pm0.02}, & \;{\rm (B)}.
\end{eqnarray}
However, there are four clusters, A119, A1367, A1656 and A2255 showing 
a large dispersion on the $\alpha$-$\beta$ plane, which  may significantly
affect the above fittings. The best-fit $\alpha$-$\beta$ relation without
these four clusters becomes 
%32,33
\begin{eqnarray}
\alpha=10^{1.16\pm0.01}\beta^{1.01\pm0.03}, & \;{\rm (A)};\\
\alpha=10^{1.41\pm0.06}\beta^{1.26\pm0.01}, & \;{\rm (B)}.
\end{eqnarray}
Alternatively, the linear fit to the data set of 45 clusters gives
%34,35
\begin{eqnarray}
\alpha=(15.40\pm2.83)\beta, & \;{\rm (A)};\\
\alpha=(15.23\pm2.54)\beta, & \;{\rm (B)}, 
\end{eqnarray}
and
%36,37
\begin{eqnarray}
r_s=(4.17\pm1.19) r_c, & \;{\rm (A)};\\
r_s=(3.70\pm1.06) r_c, & \;{\rm (B)}.
\end{eqnarray}
Within the uncertainties,
the last four relations are consistent with the findings by  
Ettori \& Fabian (1999) based on 36 high-luminosity clusters:
$\alpha=14.34\beta$ and $r_s=3.17r_c$. 

\placefigure{fig5}

\subsection{RDH vs. NFW}

We now compare the RDH and NFW profiles. Fig.6 shows  
the correlations between the density and length scale parameters 
in the two models. Applying the $\chi^2$-fit with the inclusion of   
the measurement uncertainties to the data set gives
%38,39
\begin{eqnarray}
\left(\frac{\rho_0}{\rho_c}\right)=10^{1.97\pm0.06}
     \left(\frac{\rho_s}{\rho_c}\right)^{0.73\pm0.02}, & \;{\rm (A)};\\
\left(\frac{\rho_0}{\rho_c}\right)=10^{2.18\pm0.06}
     \left(\frac{\rho_s}{\rho_c}\right)^{0.66\pm0.01}, & \;{\rm (B)},
\end{eqnarray}
and 
%40,41
\begin{eqnarray}
r_0=10^{-0.54\pm0.01}r_s^{0.71\pm0.02},  & \;{\rm (A)};\\
r_0=10^{-0.54\pm0.01}r_s^{0.64\pm0.02},  & \;{\rm (B)}.
\end{eqnarray}
The average ratios of $\rho_0/\rho_s$ and $r_0/r_s$ are, respectively,
%42,43
\begin{eqnarray}
\rho_0=(8.63\pm5.61)\rho_s, & \;{\rm (A)};\\
\rho_0=(7.89\pm4.93)\rho_s, & \;{\rm (B)},
\end{eqnarray}
and
%44,45
\begin{eqnarray}
r_0=(0.36\pm0.09) r_s, & \;{\rm (A)};\\
r_0=(0.37\pm0.10) r_s, & \;{\rm (B)}.
\end{eqnarray}
The last two relations are consistent with the estimate of Burkert (2000),
$r_0\approx 0.2r_s$, for dwarf galaxies.
The existence of these strongly positive correlations is helpful
for us to make a quantitative comparison between the two models. 
Indeed, from their predicted gas densities, and in turn their X-ray
surface brightness profiles of clusters, along with the 
correlations between the characteristic density and scale length parameters, 
we are unable to distinguish the two models as the dark halos of clusters. 
However, there is a remarkable difference, the central density.   
In fact, the characteristic density parameters, $\rho_0$ and
$\rho_s$, in these two models have very different meanings. 
$\rho_0$ represents explicitly the central density of dark matter,
while the parameter $\rho_s$ corresponds to the density nowhere 
in clusters. Note that
the large error bars in the linear $\rho_0$-$\rho_s$ relation 
(eqs.[42] and [43]) and
the average values of $\rho_s$ and  $\rho_0$ listed in Table 2 are mainly
due to the inclusion of a very few clusters (e.g. A262 and A3526)
whose X-ray surface brightness profiles show two distinct length scales

\placefigure{fig6}

\subsection{Central dark matter density}

Recall that the disagreement between the shallower central density profiles
required by various observations  with the cusped density profile 
provided by the NFW model is one of the primary motivations for advocating the
scenario of the weakly interacting dark matter particles
(Spergel \& Steinhardt 2000). A combination of the RDH profile 
and the X-ray surface brightness measurements of clusters
can now allow us to determine straightforwardly the central dark matter 
densities ($\rho_0$) of clusters. It is easy to show that
$\rho_0$ is related to the central total mass density $\rho_{\beta,0}$
inferred from the conventional isothermal $\beta$ model through
%46
\begin{equation}
\rho_{\beta,0}=\frac{9\beta kT}{4\pi G\mu m_p r_c^2}=
               9\left(\frac{\beta}{\alpha_0}\right)
	       \left(\frac{r_0}{r_c}\right)^2\rho_0.
\end{equation}
Using the linear relations established above between $\alpha_0$, $\beta$, 
$r_0$ and $r_c$, we have $\rho_{\beta,0}\approx\rho_0$.
Of course, the good agreement between $\rho_{\beta,0}$ and $\rho_0$
could be interpreted as the consequence of the common working hypothesis
behind the two models: The intracluster gas is assumed to be 
isothermal and in hydrostatic equilibrium  with the underlying gravitational
potential. In a recent work, Firmani et al. (2000) have demonstrated  
an essentially constant central density of dark halos for  
a broad range of masses from dwarf galaxies to clusters of galaxies.
However, there are only three data points on cluster scales which are 
derived from gravitational lensing. Now, we superimpose our
derived central densities of dark halos of the 45 MME clusters on  
their illustration of halo central density vs. maximum rotation velocity
(Fig.7), in which the velocity dispersion is plotted as the  
horizontal axis for clusters. Although the central densities of the clusters 
span almost two decades from 
$10^{-3}$ to $10^{-1}$ M$_{\odot}$ pc$^{-3}$,
the average value
$\langle\rho_0\rangle=0.012$ M$_{\odot}$ pc$^{-3}$ 
($0.006 M_{\odot}$ pc$^{-3}$ for model B) is in agreement with 
the one ($0.02$ M$_{\odot}$ pc$^{-3}$) reported by Firmani et al. (2000).
This reinforces the claim for the presence of the soft halo cores 
over the entire mass range.

\placefigure{fig7}

\section{Discussion and conclusions}

None of the present numerical simulations based on typical CDM models can 
reproduce the soft cores of dark halos detected  
observationally in various systems
from dwarf and low surface brightness galaxies to clusters of galaxies. 
This challenges the analytic and elegant form of the universal density 
profile suggested by NFW as the virialized dark halos. 
Without a sophisticated treatment of the
dynamical evolution of the CDM particles, we have made an attempt to
adopt the empirical
density profile (RDH) proposed by Burkert (1995) to replace the cusped NFW
profile for dark halos. The RDH profile with a finite core $r_0$ 
has been shown to give a perfect fit to the observed rotation curves of
dwarf galaxies. In particular, it also provides an excellent fit to the dark
matter distributions revealed by numerical simulations (Burkert 2000)
in the scenario that the CDM particles are weakly self-interacting, 
suggested recently
to overcome the difficulties of CDM models (Spergel \& Steinhardt 2000).
In the present paper we have applied the RDH profile to clusters of galaxies
and derived the radial distribution of intracluster gas under the
isothermal and hydrostatic equilibrium hypotheses. It turns out that
the RDH resulted gas density resembles the conventional $\beta$ model,
although a slightly large $\beta$ parameter ($\beta\sim0.7$-$0.9$)
may be required. The latter nevertheless agrees with the results usually 
found by numerical simulations (e.g. NFW; Eke, Navarro \& Frenk 1998), 
and is also marginally consistent with
the $\beta$ model fit by excising the cooling flow regions 
(e.g. Vikhlinin et al. 1999)  or adopting an additional $\beta$ model 
for the central X-ray emission (Xue \& Wu 2000).
Except for the very differently asymptotic behaviors at $r\rightarrow0$,
the RDH and NFW profiles become to be almost indistinguishable 
from each other within the extent of the current X-ray observations
of clusters. By fitting the RDH predicted X-ray surface brightness profile
to the observed one for an ensemble of 45 clusters, we have estimated the
typical central density $\rho_0$ and core radius $r_0$ in the
RDH profile, which show a fairly strong correlation with the characteristic
density $\rho_s$ and scale parameter $r_s$ in the NFW profile, respectively.
Meanwhile, our derived central dark matter densities of the 45 clusters 
have an average value of 
$\langle\rho_0\rangle\approx0.01$ M$_{\odot}$ pc$^{-3}$, 
in agreement with the result estimated on galactic scales 
(Firmani et al. 2000).

Yes, a conclusive justification for whether the RDH profile can be used as a
good approximation of dark halos and therefore, replace the role of
the NFW profile for typical CDM models will be provided 
by high-resolution simulations incorporated with new physical mechanism
for dark matter particles such as the weak self-interaction.
Several new models with the help of numerical simulations are being
constructed by a number of authors.  It can be predicted that 
a new analytic model with an isothermal core, 
if it is not the RDH profile, 
for the structure of virialized dark halos will soon be available.
The rotation curves of dwarf galaxies, the total mass distribution of 
clusters revealed by gravitational lensing, and 
the X-ray properties of clusters as explored in the present paper
will constitute a critical test for the new RDH profile.

\acknowledgments
This work was supported by 
the National Science Foundation of China, under Grant 19725311.

%\clearpage
%\setcounter{page}{1}
%\include{table-include}
\clearpage

\figcaption{The radial density profiles of intracluster gas derived 
from the RDH profiles (solid lines) and the superimposed 
$\beta$ models (open squares). 
The parameters ($\alpha_0$, $\beta$, $r_c/r_0$) 
are: (a)-(5, 0.40, 0.86), (b)-(10, 0.74, 0.85) and (c)-(20, 1.56, 0.92).
For comparison, the NFW predicted results (dotted lines)
are also shown with $\alpha=\alpha_0$.    
The horizontal scales are $r/r_0$ and $r/r_s$ for the RDH and NFW profiles,
respectively.
\label{fig1}}

\figcaption{Numerical solutions of gas density and baryon fraction,
when the self-gravity of the gas is included,   
for two choices of the central density $\delta_c=\rho_0/\rho_c$ and
the universal baryon fraction $f_{b,BBN}$. 
The best-fit $\beta$ model to each curve is plotted as open squares.
The corresponding parameters are summarized in Table 1. 
\label{fig2}}

\figcaption{An example of the observed and predicted X-ray surface
brightness profiles $S_x$ for A3158.  Filled circles: the ROSAT PSPC 
observed data; Open squares: the best-fit $\beta$ model; 
Dashed line: the NFW predicted result; Solid line: the RDH result.
Residuals between the best-fit of the RDH profile and the data are
illustrated in upper panel.  
\label{fig3}}

\figcaption{Correlations between the slope and scale parameters 
in the RDH profile and $\beta$ model determined from the 45
MME clusters. The solid lines are the best-fit relations.
\label{fig4}}

\figcaption{The same as Fig.4 but for NFW profile and $\beta$ model.
The dotted line shows the best-fit relation by excluding
A119, A1367, A1656 and A2255. 
\label{fig5}}

\figcaption{Correlations between the density and scale parameters 
in the RDH and NFW profiles derived from the 45 MME clusters.
The solid lines represent the best-fit relations.
\label{fig6}}

\figcaption{Central density of dark halos 
on scales from galaxies to clusters of galaxies. We take the data for 
dwarf galaxies (filled squares), LSB galaxies (open squares) and three distant 
clusters (open circles) from the work of Firmani et al. (2000) directly.
Our derived central densities $\rho_0$ in the RDH profile for 45 MME 
clusters are shown by filled circles, for which the horizontal axis 
represents velocity dispersion rather than maximum rotation
velocity ($V_m$)  as for galaxies.
\label{fig7}}

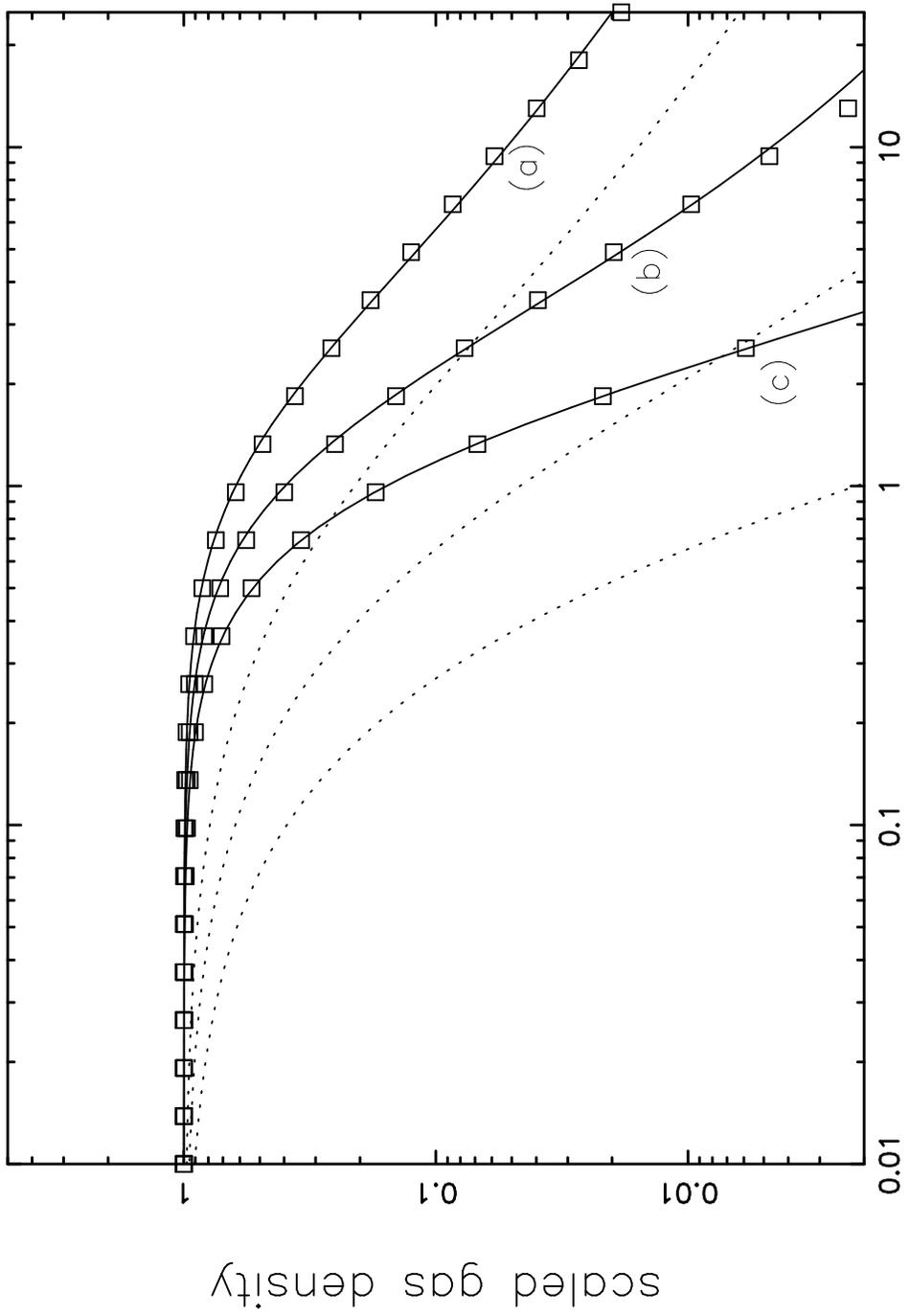
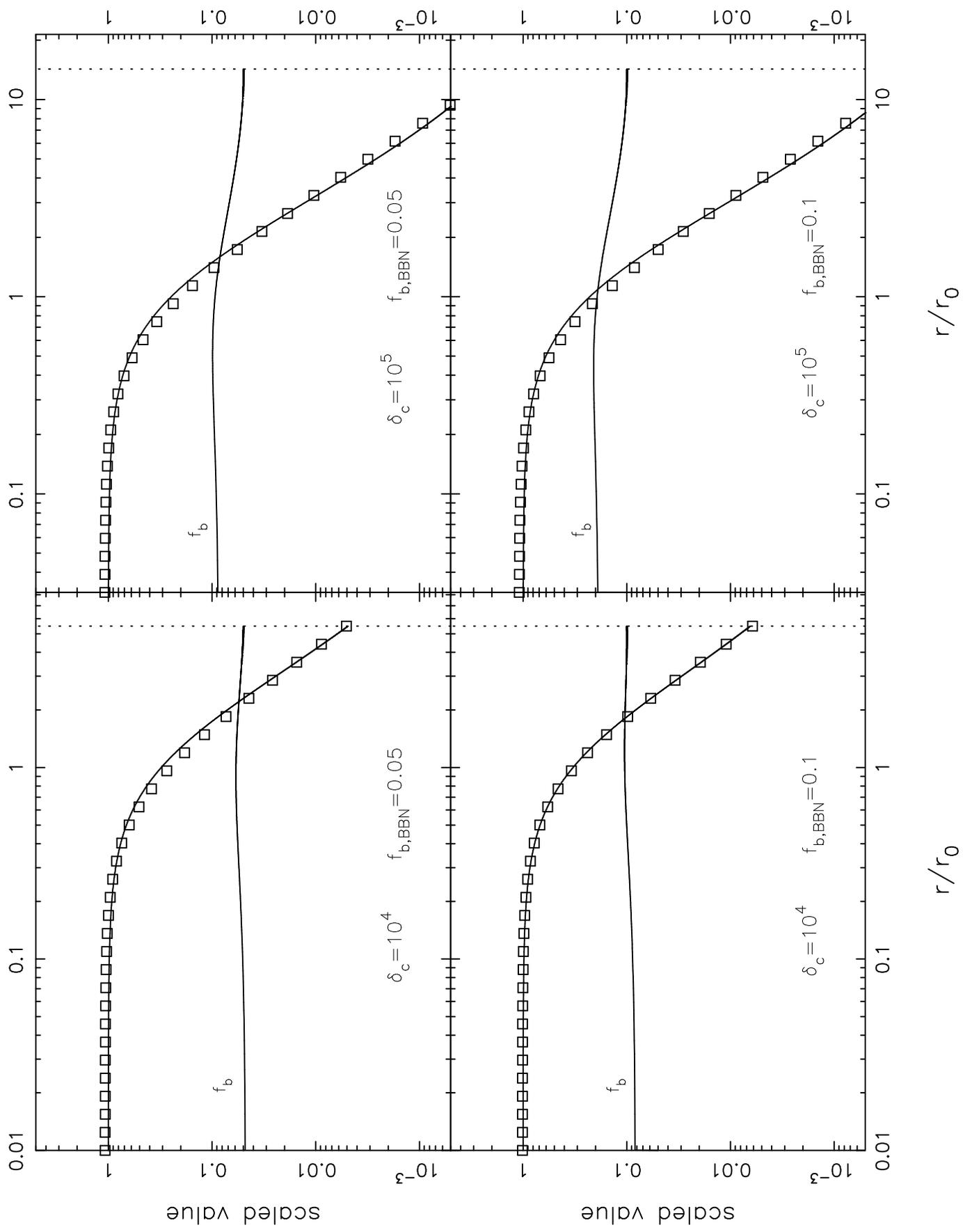
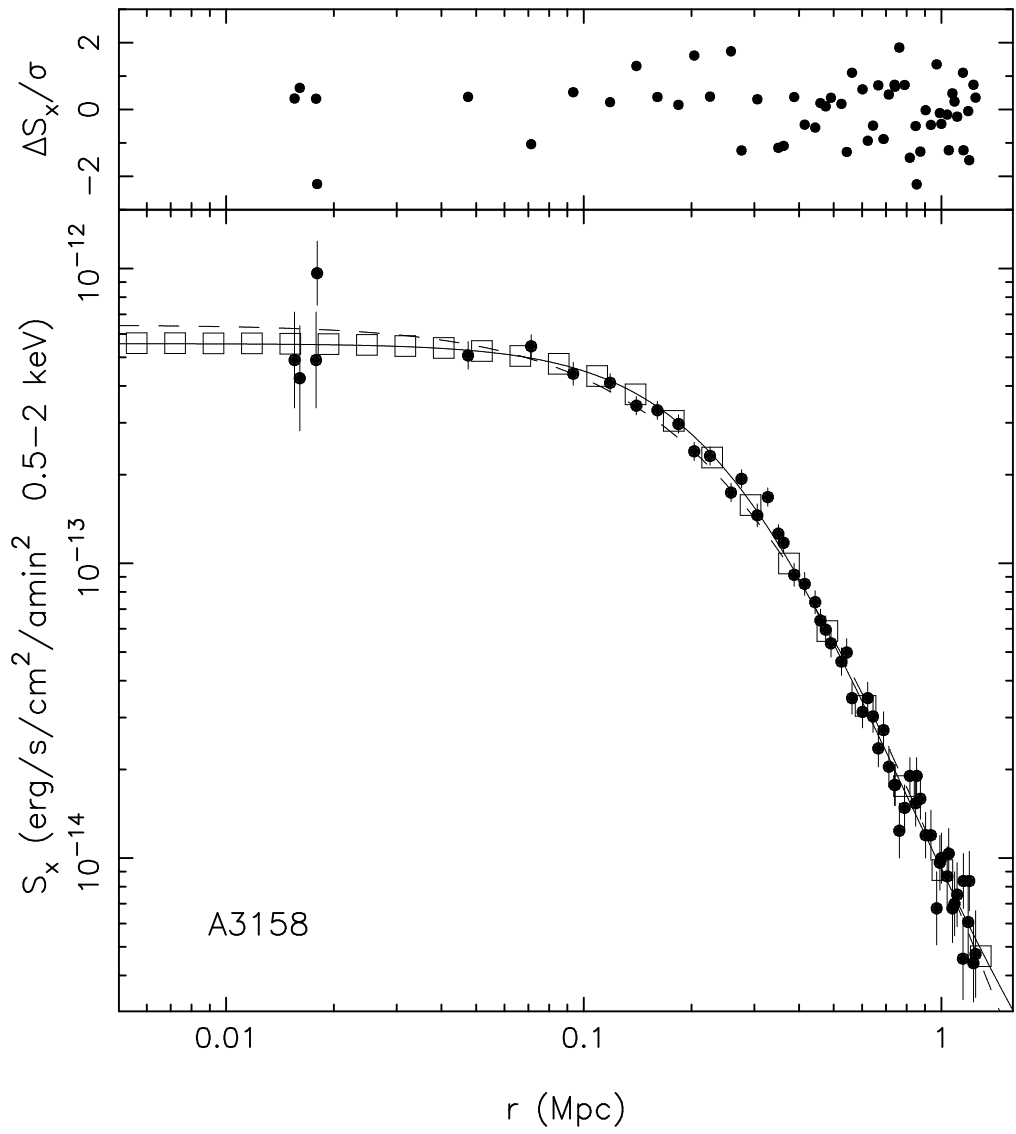
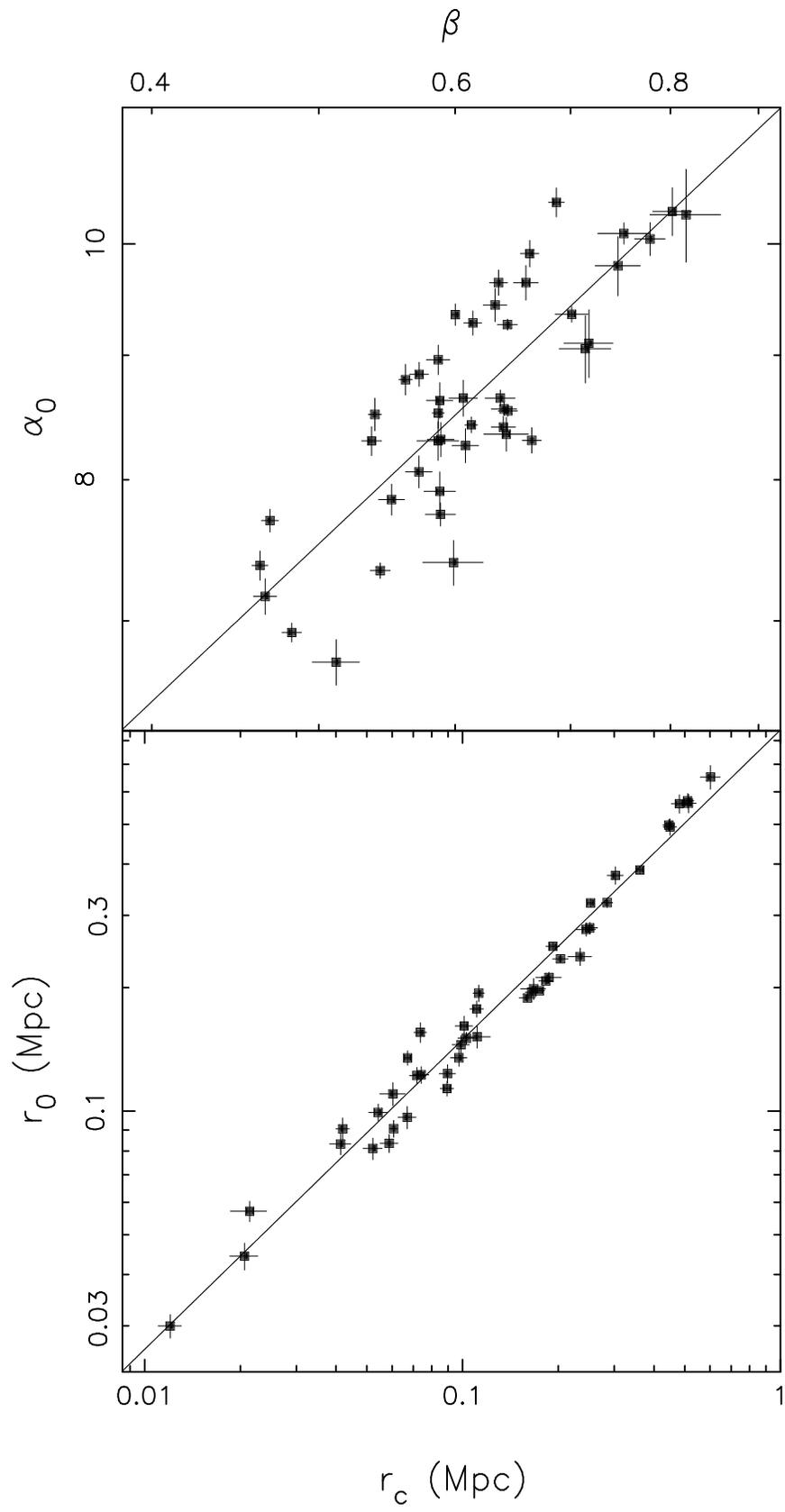
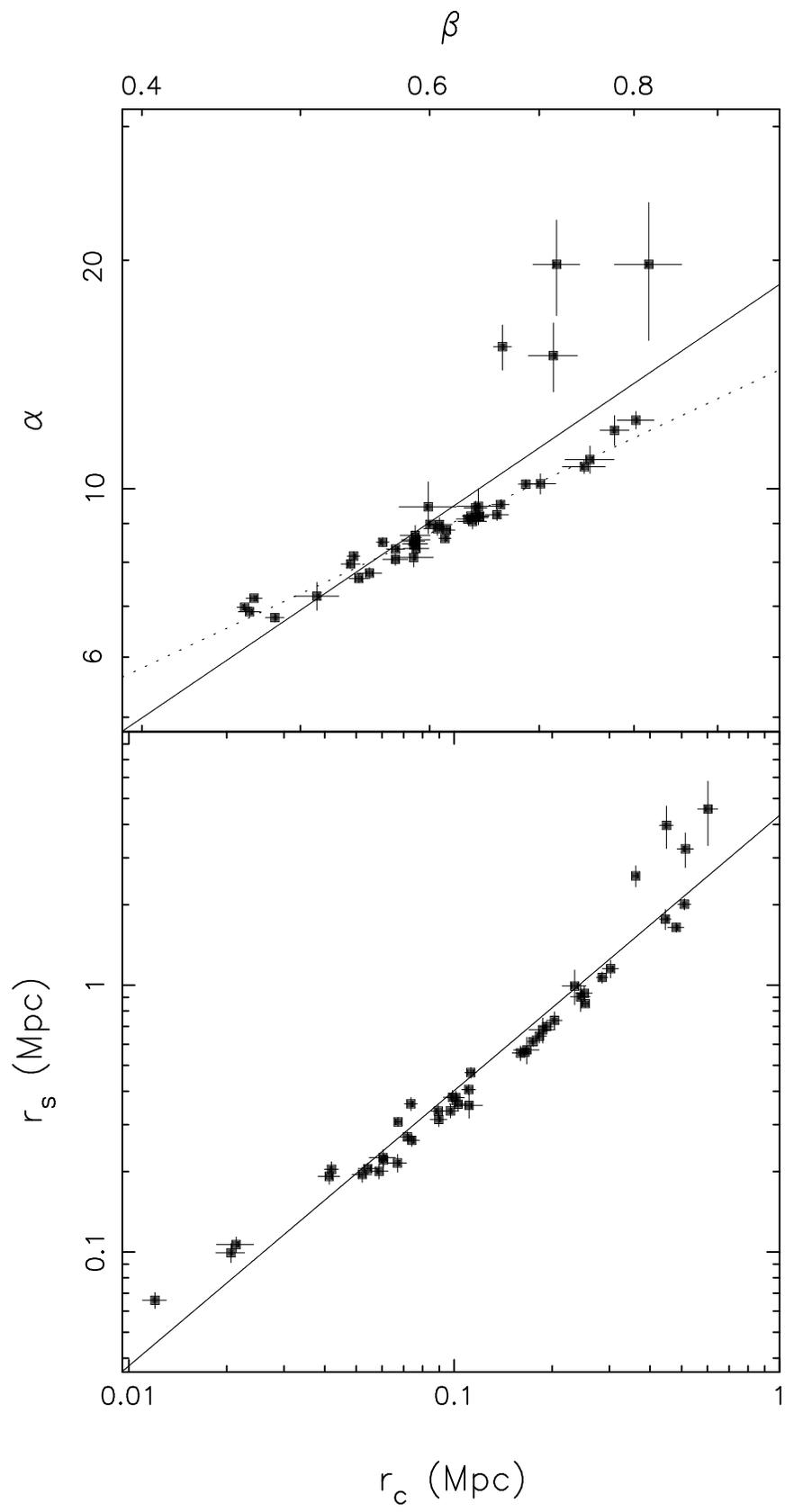
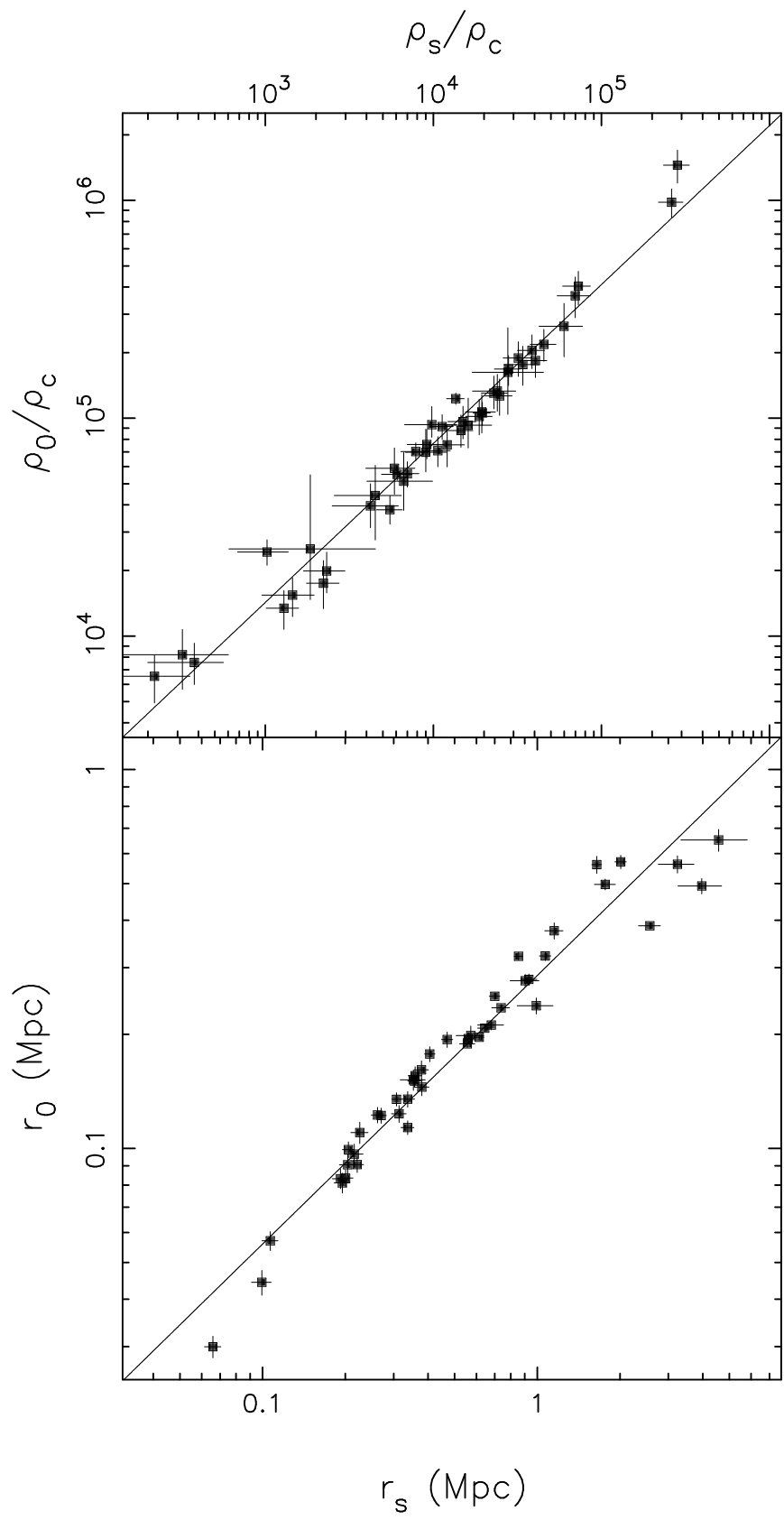
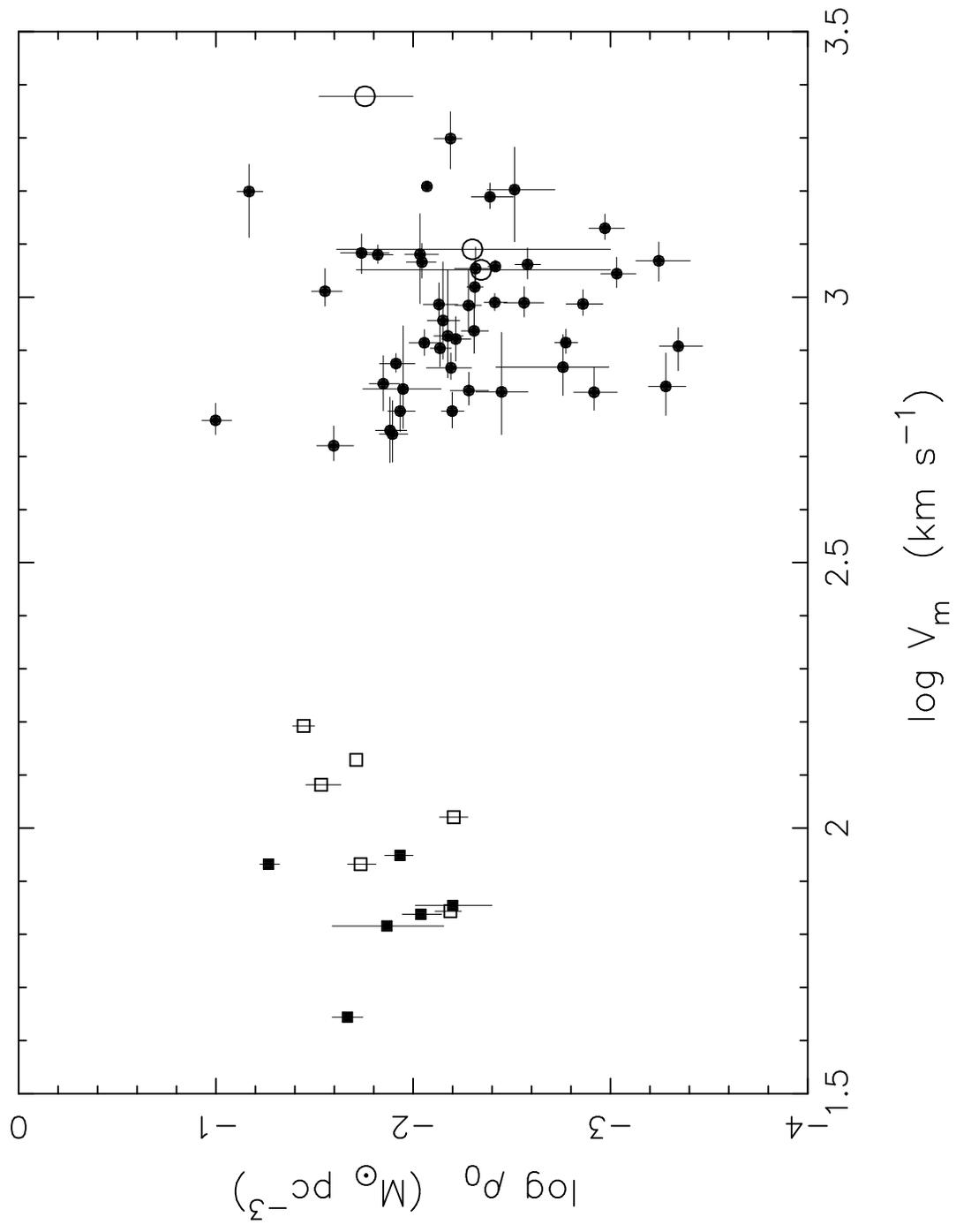
\end{document}